\documentclass{article}

\usepackage{graphicx}

\newcommand\alphaz{\alpha_z}
\newcommand\Sipi{\mbox{\rm Si}(\pi)}
\newcommand\GibbsC{C_{{\!}_{\scriptstyle G}}}
\newcommand\osc{\mathop{\mbox{\rm osc}}}

  \newcommand\CODATA
{1998 CODATA recommended values of the fundamental physical
constants}
  \newcommand\captionfigone
{Partial sum $s_n$ of the square wave for $n$=200}
  \newcommand\captionfigtwo
{Comparison of $\alphaz^{-1}$ given by (\ref{eq:result}) with the
experimental values which were used to determine $\alpha^{-1}$}

\title{An observation on the relation between the fine structure constant
and the Gibbs phenomenon in Fourier analysis}

\author{\small Zi-Xiang Zhou\\
\small Institute of Mathematics, Fudan University, Shanghai
200433, P.R.China\\ \small Email: zxzhou@guomai.sh.cn}

\begin{document}

\date{}

\maketitle

\begin{abstract}
A value given by a simple mathematical expression is proposed
which is close to the fine structure constant given by 1998 CODATA
recommended values of the fundamental physical constants up to
relative accuracy $10^{-7}$. This expression relates closely with
the value of the overshoot of the Gibbs phenomenon in Fourier
analysis.
\end{abstract}



\section{Introduction}

The fine structure constant $\alpha$ is the most important
dimensionless universal physical constant. Since it is
dimensionless and universal, it is very interesting to know
whether it can be expressed by a simple expression of universal
mathematical constants. This may help understanding the nature.

As is known, the definition of fine structure constant $\alpha$ is
\begin{equation}
   \alpha^{-1}=\frac{2c\epsilon_0h}{e^2}=\frac{2h}{c\mu_0e^2}
\end{equation}
where $c$ is the speed of light in vacuum, $\mu_0$ and
$\epsilon_0=1/c^2\mu_0$ are the permeability and permittivity of
vacuum respectively, $e$ is the elementary charge and $h$ is the
Planck constant.

Here we propose
\begin{equation}
   \alphaz^{-1}=\frac 1{\sqrt{2}}\left(\frac{3\pi}2\right)^3
    \int_0^\pi\frac{\sin x}{x}\,\mbox{\rm d} x
    \approx 137.03598260.
   \label{eq:result}
\end{equation}
This value coincides with the value
\begin{equation}
   \alpha^{-1}=137.03599976\pm 0.00000050
\end{equation}
given by the \CODATA{} \cite{CODATA1998} up to relative accuracy
$1.26\times 10^{-7}$.

In the expression (\ref{eq:result}), the integral
$\displaystyle\Sipi=\int_0^\pi\frac{\sin x}x\,\mbox{\rm d} x$ is
related to a universal mathematical constant --- the overshoot of
the Gibbs phenomenon in Fourier analysis, which is the same to all
the jump discontinuities of all piecewisely smooth functions. The
expression $\alphaz^{-1}/\Sipi=(3\pi/2)^3/\sqrt{2}$ is so simple
that it seems that there may be physical essence behind the
relation $\alphaz\approx \alpha$, although it is not known
presently.

In the next section, the universality of the constant $\Sipi$ in
Fourier analysis is reviewed briefly. Then, the value $\alphaz$ is
compared with the values of $\alpha$ given by experiments and
physical theories, which were used to determine $\alpha$ for the
\CODATA. At last, some discussions are presented.

\section{A brief review of the Gibbs phenomenon}

The Gibbs phenomenon is a universal phenomenon at all the jump
discontinuities for Fourier series or Fourier transformations.
First, let us look at a simple example. Let $f(x)$ be a square
wave of period $2\pi$ with $f(x)=1$ for $0<x<\pi$ and $f(x)=-1$
for $-\pi<x<0$. Let $s_n$ be the partial sum of its Fourier
series, i.e.
\begin{equation}
   s_n(x)=\frac 4{\pi}\sum_{k=1}^n\frac{\sin(2k-1)x}{2k-1}.
\end{equation}
Then when $n\to\infty$, there are overshoots and undershoots in
the graph of $s_n$ (Fig.~1).

\begin{figure}\begin{center}
\includegraphics{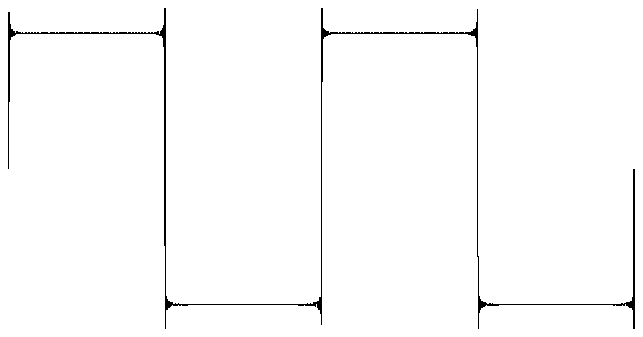} \caption{\captionfigone}
\end{center}\end{figure}

The limit of the amplitude of $s_n$ near $0$ as $n\to\infty$ is
$2\GibbsC$, which is $\GibbsC$ times the jump of $f(x)$ at $0$.
Here
\begin{equation}
   \GibbsC=\frac 2{\pi}\Sipi=\frac{2}{\pi}\int_0^\pi
   \frac{\sin x}{x}\,\mbox{\rm d} x
   \approx 1.17897974447217.
   \label{eq:cG}
\end{equation}
This is the famous Gibbs phenomenon
\cite{Hewitt,Korner,Walker,Champeney}, which was discovered more
than a century ago.

The Gibbs phenomenon appears not only in the square wave, but also
at all jump discontinuities generally. For the Fourier series of
any piecewisely smooth period function $f(x)$, the limit of the
amplitude of the partial sum at a jump discontinuity $x_0$ of
$f(x)$ equals to $\GibbsC$ times the jump of $f(x)$ at that point.
That is, for the partial sum $s_n(x)$,
\begin{equation}
   \lim_{\delta\to 0^+}\lim_{n\to\infty}\osc_{|x-x_0|\le
    \delta}s_n(x)\nonumber
   =\GibbsC\lim_{\delta\to 0^+}\osc_{|x-x_0|\le
    \delta}\lim_{n\to\infty}s_n(x)
\end{equation}
where {\it osc} refers to the difference of the maximum and the
minimum of a function. This fact is also true for the Fourier
transformation of a non-periodic function, if the function is
absolutely integrable and piecewisely smooth \cite{Champeney}. In
all cases, the constant $\GibbsC$ is the same.

\section{Comparison with experimental values}

In Fig.~2, $\alphaz^{-1}$ given by (\ref{eq:result}) was compared
with the values of $\alpha^{-1}$ given by the experimental data
together with various physical theories. Each line segment in the
figure represents a datum in Table XV of \cite{CODATA1998}. That
table was used to determine the value $\alpha$ for the \CODATA.
For each value $x$ with uncertainty $\sigma$, the line segment
extends from $x-\sigma$ to $x+\sigma$ in Fig.~2. The
identifications to the right of the line segments are the same as
those in Table XV of \cite{CODATA1998}. The vertical line marked
$\alphaz^{-1}$ represents the value given by (\ref{eq:result}) and
that marked $\alpha^{-1}$ represents the value given by the
\CODATA.

\begin{figure}\begin{center}
\includegraphics{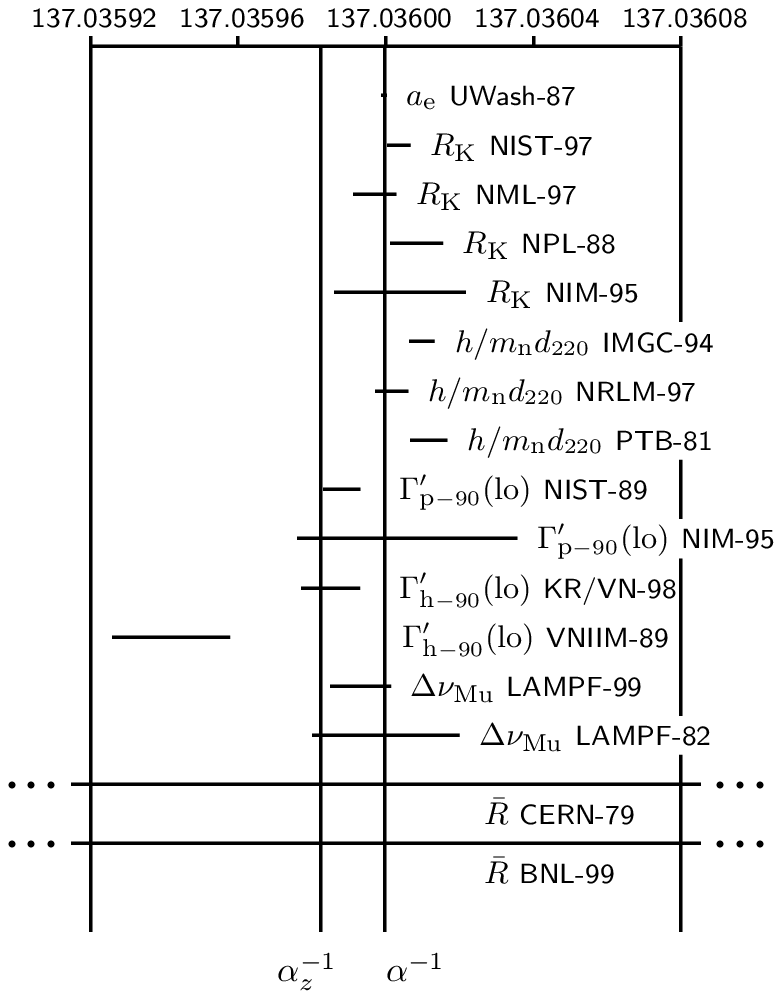} \caption{\captionfigtwo}
\end{center}\end{figure}

\section{Discussions}

There are other theoretical values of $\alpha$, such as those
given by the string theory \cite{Naschie1,Naschie2}. For example,
the simplest expression is
$\alpha^{-1}=\phi^{-10-\phi^3}=137.7880938$ where
$\phi=(\sqrt5-1)/2$ is the golden mean. More accurate values with
accuracy $3\times 10^{-7}$ were also given by a complicated
expansion \cite{Naschie2}. Although the expression $\alphaz^{-1}$
in this paper is simply an observation, it is much simpler than
the other theoretical results with similar accuracy.

There is a small possibility that $\alphaz^{-1}\approx\alpha^{-1}$
is simply an accidental coincidence. Actually it is not difficult
to construct complicated relations of $2,3,\pi$ etc. to
approximate $\alpha^{-1}$ because any real number can be
approximated by a rational number to any accuracy. For example,
$2^{-19/6}3^{157/24}\pi^{-1/16}=137.0360046$. (There is also a
number $2^{19/4}3^{-7/4}5^{1/4}\pi^{11/4}=137.036082$ given by
Wyler \cite{Naschie1}.) However, the probability is very small to
get accidentally a simple relation like
$\alphaz^{-1}/\Sipi=(3\pi/2)^3/\sqrt{2}$ in this paper. The
internal relation between $\alphaz$ and the fine structure
constant is to be revealed.

\end{document}